\documentclass{cupconf}

  \checkfont{eurm10}
  \iffontfound
    \IfFileExists{upmath.sty}
      {\typeout{^^JFound AMS Euler Roman fonts on the system,
                   using the 'upmath' package.^^J}%
       \usepackage{upmath}}
      {\typeout{^^JFound AMS Euler Roman fonts on the system, but you
                   dont seem to have the}%
       \typeout{'upmath' package installed. cupconf.cls can take advantage
                 of these fonts,^^Jif you use 'upmath' package.^^J}%
      }
  \else
  \fi

  \checkfont{msam10}
  \iffontfound
    \IfFileExists{amssymb.sty}
      {\typeout{^^JFound AMS Symbol fonts on the system, using the
                'amssymb' package.^^J}%
       \usepackage{amssymb}%
       \let\le=\leqslant  
       \let\ge=\geqslant  
      }{}
  \fi

  \IfFileExists{amsbsy.sty}
    {\typeout{^^JFound the 'amsbsy' package on the system, using it.^^J}%
     \usepackage{amsbsy}}
    {\providecommand\boldsymbol[1]{\mbox{\boldmath $##1$}}}

\newsavebox{\astrutbox}
\sbox{\astrutbox}{\rule[-5pt]{0pt}{20pt}}

\newcommand\ga{\stackrel{_>}{_\sim}}
\newcommand\la{\stackrel{_<}{_\sim}}
\newcommand{\kms}{km\,s$^{-1}$}

\title[Hot Gas in the Local Universe]{Hot Gas in the Local Group and Low-Redshift 
Intergalactic Medium}

\author[K. R. Sembach]%
{K\ls E\ls N\ls N\ls E\ls T\ls H\ns R.\ns S\ls E\ls M\ls B\ls 
A\ls C\ls H\ns}

\affiliation{$^1$The Space Telescope Science Institute, 3700 San Martin
Dr., Baltimore, MD  21218, USA\\[\affilskip]}

\pubyear{}
\volume{}
\pagerange{}
\date{?? and in revised form ??}
\begin{document}

\maketitle

\begin{abstract}
There is increasing observational 
evidence that hot, highly ionized interstellar and 
intergalactic gas plays a significant role in the evolution of galaxies in 
the local universe. 
The primary spectral diagnostics of the warm-hot 
interstellar/intergalactic medium are ultraviolet and X-ray absorption 
lines of O\,{\sc vi} and O\,{\sc vii}. 
In this 
paper, I summarize some of the recent highlights of spectroscopic studies 
of hot gas in the Local Group and low-redshift universe. 
 These highlights include 
investigations of the baryonic content of low-$z$ O\,{\sc vi} absorbers,
evidence for a hot Galactic corona or Local Group medium, and the 
discovery of a highly ionized high velocity cloud system around the Milky Way.
\end{abstract}

\firstsection % if your document starts with a section,
\section{Introduction}

We live in a wonderful age of discovery and exploration of the universe.
As we peer farther and farther back in time, it is becoming ever more 
important to make sure that we observe the local universe as well
as possible.  Observations of galactic systems and the intergalactic 
medium (IGM) in the low-redshift universe are required to study the universe
as it has evolved over the last $\sim5$ billion years.   They are 
essential for the interpretation of higher redshift systems, and they
form a framework for studies of such key topics as galactic evolution,
``missing mass'', and the distribution of dark matter.  Studies of 
hot gas and its relationship to galaxies are shedding new light 
on these and other astronomical topics of interest today.
In this review, I summarize some basic information about the elemental 
species and types of observations that can be used to study hot gas.  
I also  provide overviews of recent spectroscopic observations of 
hot gas in the low-redshift universe (\S2), the Local Group (\S3), and
the high velocity cloud (HVC) system that surrounds the Milky Way (\S4).
Some concluding remarks on future observations can be found in \S5.

Table~1 contains a summary of some of the most important diagnostics of 
hot gas in the local universe.  The species listed are detectable in 
either the ultraviolet or X-ray bandpasses accessible with spectrographs 
aboard the Hubble Space Telescope (HST), the Far Ultraviolet Spectroscopic 
Explorer (FUSE), the Chandra X-ray Observatory, and XMM-Newton.  Most of
the diagnostics listed can be observed at their rest wavelength, which 
means that they can be used to study hot gas in the Local Group.  Others
must be redshifted into one of the observable bandpasses; the 
observed wavelength at $z=0.5$ is listed for for comparison in the table.

\begin{table}
\begin{center}
\caption{Diagnostics of Hot Gas at Low Redshift ($z < 0.5)^a$}
\begin{tabular}{lccccc}
\hline
Ion, $\lambda_{\rm rest}$ & $f$  & $\log f\lambda$  & $\lambda_{z=0.5}$ 
& $T_{CIE}$$^b$ 
& b$_{th}$$^c$ \\
& & & (\AA) & (K) & (km~s$^{-1}$)\\
\hline
\multicolumn{6}{c}{FUSE, HST}\\
\hline
H\,{\sc i} Ly-series & \ldots& \ldots & 1369-1824  & \ldots & 13--129$^d$\\
C\,{\sc iv} 1548.195  & 0.1908 & 2.470 & 2322.292 & $1.0\times10^5$ & 11.8 \\
C\,{\sc iv} 1550.770  & 0.0952 & 2.169 & 2326.155 & $1.0\times10^5$ & 11.8 \\
N\,{\sc v} 1238.821 &  0.1570 & 2.289 & 1858.232 & $1.8\times10^5$ & 14.6 \\
N\,{\sc v} 1242.804 & 0.0782 & 1.988 & 1864.206 & $1.8\times10^5$ & 14.6 \\
O\,{\sc iv} 787.711 & 0.111 & 1.942 & 1181.567 & $1.6\times10^5$ & 12.9 \\
O\,{\sc v} 629.730 & 0.515 & 2.511 & \phantom{0}944.595 & $2.5\times10^5$ & 16.1 \\
O\,{\sc vi} 1031.926 & 0.1329 & 2.137 & 1547.889 & $2.8\times10^5$ & 17.1 \\
O\,{\sc vi} 1037.617 & 0.0661 & 1.836 & 1556.425 & $2.8\times10^5$ & 17.1 \\
Ne\,{\sc viii} 770.409 & 0.103 & 1.900 & 1155.614 & $5.6\times10^5$ & 21.5 \\
Ne\,{\sc viii} 780.324 & 0.0505 & 1.596 & 1170.486 & $5.6\times10^5$ & 21.5 \\
\hline
\multicolumn{6}{c}{Chandra, XMM-Newton} \\
\hline
O\,{\sc vii} 21.602 & 0.696 & 1.177 & \phantom{0}\phantom{0}32.403 & $8.0\times10^5$ & 28.8 \\
O\,{\sc viii} 18.967 & 0.277 & 0.720 & \phantom{0}\phantom{0}28.450 & $2.2\times10^6$ & 47.8 \\
O\,{\sc viii} 18.972 & 0.139 & 0.421 & \phantom{0}\phantom{0}28.459 & $2.2\times10^6$ & 47.8 \\
Ne\,{\sc ix} 13.447 & 0.0724 & 0.988 & \phantom{0}\phantom{0}20.170 & $1.5\times10^6$& 35.2 \\
\hline
\end{tabular}
\end{center}
$^a${$f$-values and wavelengths (in \AA) are from Morton (1991),
Verner, Barthel, \& Tytler (1994), and Verner, Verner, \& Ferland (1996).}

$^b${Temperature of maximum ionization fraction in 
collisional ionization equilibrium (Sutherland \& Dopita 1993).}

$^c${Thermal line width, b = $(2kT/m)^{1/2}$, at 
$T=T_{CIE}$ unless indicated otherwise.}

$^d${Value of b for $T=10^4-10^6$\,K.}

\end{table}

The O\,{\sc vi} $\lambda\lambda1031.926, 1037.617$ resonance doublet lines
are the best lines to use for 
kinematical investigations of hot ($T \sim 10^5-10^6$\,K) gas in the 
low-redshift universe.  O\,{\sc vi} has a higher ionization potential
than other species observable by FUSE and HST, and oxygen has the
highest cosmic abundance of all elements other than hydrogen and 
helium.
X-ray spectroscopy of the interstellar or intergalactic gas in 
higher ionization lines (e.g., O\,{\sc vii}, O\,{\sc viii}) is possible with
XMM-Newton and the Chandra X-ray Observatory for a small number 
of sight lines toward AGNs and QSOs, but the spectral resolution 
(R~$\equiv \lambda/\Delta\lambda \la 400$)
is modest compared to that afforded by FUSE (R~$\sim 15,000$) or HST/STIS
($R\sim45,000$).  While 
the X-ray lines provide extremely useful information about the amount of 
gas at temperatures greater than $10^6$\,K, the interpretation of where that
gas is located, or how it is related to the $10^5-10^6$\,K gas traced by
O\,{\sc vi}, is hampered at low redshift by the kinematical
complexity of the hot interstellar medium (ISM) and 
IGM along the sight lines observed, as discussed 
below.  Nevertheless, 
the X-ray diagnostics are of fundamental importance for determining the 
ionization state of the nearby hot gas, and they provide information about 
hotter gas that is not traced by species in the ultraviolet wavelength 
region of the electromagnetic spectrum.

Table~2 contains a high-level summary of key considerations for 
absorption and emission-line spectroscopy of hot interstellar and 
intergalactic plasmas.  Both types of observations have their strengths,
and the combination of ultraviolet and X-ray information holds
great promise for studies of the ISM and IGM.

\begin{table}
\begin{center}
\caption{Spectroscopy of Hot Interstellar/Intergalactic Gas}
\begin{tabular}{cccc}
\hline
& {\it UV Absorption} & {\it X-ray Absorption} & {\it X-ray Emission} \\
\hline
{\it Observatory}           & FUSE, HST    & $\begin{array}{c}{\rm Chandra} \\ 
{\rm XMM-Newton} \end{array}$              & $\begin{array}{c}{\rm Chandra, ROSAT}
 \\ {\rm XMM-Newton} \end{array}$\\
\vspace{0.2cm}
$\begin{array}{c}{T_{CIE}\,{(K)}} \\ {Range} \end{array}$
& $10^5-10^6$  & $10^6-10^8$          & $10^6-10^8$\\
\vspace{0.2cm}
$\begin{array}{c}{Density} \\ {Dependence} \end{array}$         
& N$_{\rm ion} \propto n_e$  & N$_{\rm ion} \propto n_e$ & 
I$_{\rm xray} \propto n_e^2$\\
\vspace{0.2cm}
$\begin{array}{c}{Limiting} \\ {Sensitivity}^a \end{array}$ 
& $\begin{array}{c}{\rm log N(O^{+5}) \sim 13} \\ {\Rightarrow \rm log 
N(H^+) \sim 18} \end{array}$ 
& $\begin{array}{c}{\rm log N(O^{+6}) \sim 16} \\ {\Rightarrow \rm log 
N(H^+) \sim 20.3} \end{array}$ & \ldots \\
\vspace{0.2cm}
{\it Spatial Info.}  & Point source & Point source & Extended source\\
\vspace{0.2cm}
$\begin{array}{c}{Spectral} \\ {Resolution} \end{array}$
& $\begin{array}{c}{\rm R > 15,000} \\ {\rm Detailed\,kinematics} \end{array}$
& $\begin{array}{c}{\rm R < 1000} \\ {\rm General\,kinematics} \end{array}$
& $\begin{array}{c}{\rm Low\,(broadband)} \\ {\rm R < 1000} \end{array}$ \\
\hline
\end{tabular}
\end{center}
$^a${Assuming a metallicity of $\sim0.1$ solar and peak ionization
fractions (see text).}

\end{table}

\section{Low-Redshift O\,{\sc vi} Absorption Systems}

One of the recent successes of observational cosmology is the excellent 
agreement in estimates of the amount of matter contained in baryons
derived from measures of the temperature fluctuations in the 
cosmic microwave background and 
spectroscopic measures of 
the primordial
abundance of deuterium relative to hydrogen in the high-redshift intergalactic
medium  (Spergel et al. 2003). At high redshift ($z \ga 3$), 
most of the baryons in the universe are contained
in intergalactic absorbers, 
which are detected through spectroscopic observations
of their H\,{\sc i} and He\,{\sc ii} Ly$\alpha$ absorption against the background 
light of distant quasars.  At lower redshifts, the Ly$\alpha$ forest thins 
out and galaxies become more prevalent (Penton, Shull, \& Stocke 2000).  
Censuses of these 
low-redshift Ly$\alpha$ clouds and galaxies 
reveal a baryon deficit compared to the amount of matter observed 
in the high-redshift universe (see Fukugita, Hogan, \& Peebles 1998).  
This ``missing baryon'' problem has led to 
the suggestion that much of the baryonic material at low redshift is
found in the form of  hot, 
highly ionized intergalactic gas that is difficult  to detect 
with existing instrumentation.

Hydrodynamical simulations of the evolution of the IGM in the 
presence of cold dark matter predict that the intergalactic clouds collapse
into coherent sheets and filaments arranged in a web-like pattern.  As lower
density gas streams into the deeper
potential wells at the intersections of these sheets and filaments, clusters 
of galaxies form.  Shocks heat the collapsing structures to temperatures of 
$10^5-10^7$\,K, resulting in a pervasive network of hot gas
 (Cen \& Ostriker 1999;
Dav\'e et al. 2001).  Understanding the 
physical processes involved in galaxy-formation is the key to
determining whether this description of the intergalactic medium is
correct.  In particular, the  simulations do not yet have sufficient
observational constraints to accurately model 
how the cosmic web responds to the formation of galaxies.  Feedback 
between galaxies and the cosmic web affects the kinematics, distribution, 
metal content, and temperature of the IGM.  These processes are probably 
self-regulating.  Detailed studies are 
most promising at low redshift, where it is
possible to sample the IGM absorption on finer scales, identify faint 
galaxies, and study galactic properties in 
greater detail than is possible at high redshift.
Many fundamental questions remain to be answered:
\begin{enumerate}
\item{Is the hot IGM a significant repository of baryons at low redshift?  How 
much mass is contained in the cosmic web of hot gas?}
\item{How does feedback during galaxy formation affect the properties of 
the IGM and the efficiency of galaxy formation?  What are the primary feedback 
mechanisms?}
\item{What is the morphology of the cosmic web and the dark matter it 
traces?}
\item{Are there fundamental relationships 
between the hot IGM and the hot gas found in clusters and groups of galaxies?} 
\end{enumerate}

Observational evidence for a hot IGM at low redshifts remains limited mainly
to detections of highly ionized oxygen (primarily O\,{\sc vi}) along a
 small number of sight lines.  The 
low-redshift O\,{\sc vi} systems that have been observed have  
a variety of strengths and O\,{\sc vi}/H\,{\sc i} ratios. Many 
occur at the redshifts of groups or clusters of galaxies along the
same sight lines.  Several sight
lines have been examined in detail for O\,{\sc vi}
absorbers: H\,1821+643 (Tripp, Savage, \& Jenkins 2000; Oegerle et al. 2000),
PG\,0953+415 (Tripp \& Savage 2000; Savage et al. 2002), 3C\,273 
(Sembach et al. 2001), PKS\,2155-304 (Shull, Tumlinson, \& Giroux 2003), 
PG\,1259+593 (Richter et al. 2004), PG\,1116+215 (Sembach et al.
2004), and PG\,1211+143 (Tumlinson et al. 2004).
An important piece of information that is lacking for many of the 
well-observed O\,{\sc vi} absorbers is the amount of C\,{\sc iv} 
associated with the O\,{\sc vi}.  This information can, and should,
be obtained with the HST.

Figure~1 illustrates a few of the absorption lines observed by FUSE in 
one of the two O\,{\sc vi} systems toward PG\,0953+415 (Savage et al. 2002).  
The lines are 
easily detected and are nearly resolved at the FUSE resolution of 
$\sim20$ \kms\ (FWHM).  This absorption 
system has ionization properties consistent
with photoionization by  dilute ultraviolet background radiation but may 
also be collisionally ionized.
A good example of a system that is probably collisionally ionized is 
given by Tripp et al. (2001).  Other investigators have suggested that 
many of the O\,{\sc vi} systems are collisionally ionized based on the
observed relationship between the line width and O\,{\sc vi} column
density found for a wide range of environments containing O\,{\sc vi}
(Heckman et al. 2002).

\begin{figure}
\includegraphics{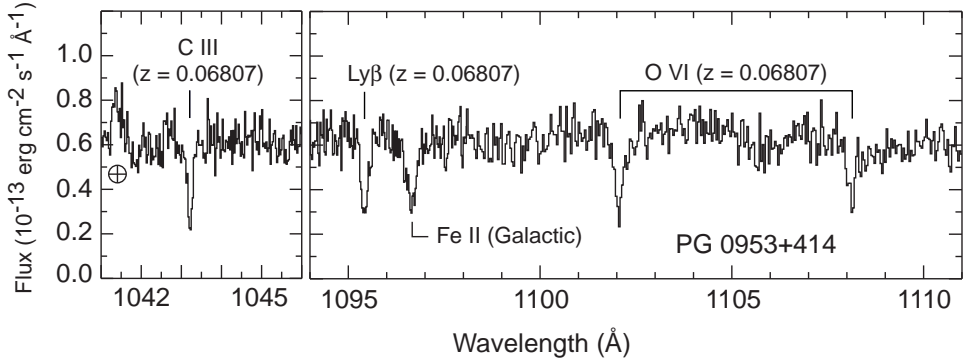}
\vspace{1.9in}
\caption{\small  A FUSE spectrum of PG\,0953+415 showing redshifted O\,{\sc vi},
C\,{\sc iii}, and H\,{\sc i} Ly$\beta$ at $z =0.06807$ (Savage et al. 2002).  
This is one of two O\,{\sc vi} systems along this sight line. 
The crossed circle in the left panel marks the location of a terrestrial 
O\,{\sc i} airglow line.}
\end{figure}

The baryonic contribution of the O\,{\sc vi} absorbers to the closure
density of the universe is $\Omega_b$(O\,{\sc vi}) = 0.002 $h_{75}^{-1}$
if their typical metallicity is $\sim1/10$ solar (Savage et al. 2002). 
Here, $h_{75}^{-1}$ is the Hubble constant in units of 75 \kms\,Mpc$^{-1}$.
This baryonic 
contribution is of the same order of magnitude as the contribution
from stars and gas inside galaxies.  More information about O\,{\sc vi} 
in the low-redshift IGM can be found in the aforementioned sight line 
references.  See Tripp (2002) for a recent review of the subject.

Information on hot gas at low redshift is also provided by 
X-ray measurements of O\,{\sc vii} and O\,{\sc viii}, although 
the number of measurements is limited at this time.  These
results are sometimes conflicting, as in the case of the $z \sim 0.057$ 
absorbers toward PKS\,2155-304 (e.g., Fang et al. 2002; Shull et al. 2003), 
or are of modest statistical significance (e.g., Cagnoni 2002;
Mathur, Weinberg, \& Chen
2003; McKernan et al. 2003b).  None of the claimed detections 
to date at $z > 0$ have been 
particularly convincing.  Still, the fact that it is possible to 
conduct searches for X-ray absorption lines in the hot IGM is promising.
To form 
a large enough database  to make firmer statements
about the baryonic contribution of gas hotter than that contained 
in the O\,{\sc vi} absorption systems will likely require spectrographs 
on X-ray telescopes with very large effective areas, such as Constellation-X.

\section{Hot Local Group Gas}

An interesting development in studies of the gaseous content of Local
Group galaxies is the mounting evidence for a hot, extended corona 
of low density gas around the Milky Way.  This corona may extend to great distances 
from the Galaxy, and may even pervade much of the Local Group.  In recent years, several
results have strengthened the case for such a medium.  These include:

\begin{enumerate}
\renewcommand{\theenumi}{\roman{enumi}}
\item{{\it The detection of {\rm O\,{\sc vi}} high velocity clouds in the vicinity of the 
Galaxy}. The properties of these clouds suggest that the O\,{\sc vi} is created at the 
boundaries of cooler HVCs as they move through a pervasive coronal medium
(Sembach et al. 2003; Collins, Shull, \& Giroux 2004). The properties of the 
O\,{\sc vi} HVCs and their possible origins are discussed below in \S4. }

\item{{\it Chandra and XMM-Newton detections of {\rm O\,{\sc vii}}
 absorption near 
zero redshift.}  O\,{\sc vii} peaks in abundance at $T \sim 10^6$\,K in collisional 
ionization equilibrium.  A summary of these O\,{\sc vii}
detections is given in Table~3, where the observed equivalent width of the 
O\,{\sc vii} $\lambda21.60$ line is listed.  Other high ionization species, such as
O\,{\sc viii} $\lambda18.97$ and Ne\,{\sc ix} $\lambda13.45$ may 
also be detected along the same sight lines, albeit at somewhat lower significance. }

\item{{\it Ram-pressure stripping of gas in Local Group dwarf galaxies.}  
 Moore \& Davis (1994) postulated a hot,
low-density corona to provide ram pressure stripping of some of the Magellanic
Cloud gas and to explain the absence of gas in globular clusters and 
nearby dwarf spheroidal companions to the Milky Way (see also Blitz \&
Robishaw 2000).  The shape and confinement 
of some Magellanic Stream concentrations
(Stanimirovic et al. 2002) and the shapes of supergiant shells along
the outer edge of the LMC (de~Boer et al. 1998) are also more easily explained
if an external medium is present.}

\item{{\it Drag deflection of the Magellanic Stream on its orbit around the 
Milky Way.} N-body simulations
of the tidal evolution and structure 
of the leading arm of the 
Magellanic Stream require a low-density medium ($n_H < 10^{-4}$ cm$^{-3}$)
to deflect some of the Stream gas into its observed configuration (Gardiner 1999). }

\end{enumerate}

\begin{table}
\begin{center}
\caption{Reported Detections of O\,{\sc vii} $\lambda 21.60$ Absorption Near $z\sim0$}
\begin{tabular}{lccccc}
\hline
Sight Line      & l\,($^o$) & b\,($^o$) & $W_\lambda$(m\AA)$^a$ & Instrument & References$^b$ \\
\hline
PKS\,2155-304   & \phantom{0}17.73 & --52.25 & $11.6\pm^{7.5}_{6.0}$ & Chandra LETG & N02,F03 \\
		&	&	& $15.6\pm^{8.6}_{4.9}$ & Chandra LETG & F02\\
		& 	& 	 & $16.3\pm3.3$ & XMM RGS & R03 \\
\\
Mrk\,421        & 179.83 & +65.03 &  $15.4\pm1.7$ & XMM RGS & R03 \\
\\
3C\,273         & 289.95 & +64.36 &$28.4\pm^{12.5}_{\phantom{0}6.2}$ & Chandra LETG & F03 \\
                &        &       &$26.3\pm4.5$            & XMM RGS     & R03 \\
\\
NGC\,4593       & 297.48 & +57.40 &  $18.0\pm^{\phantom{0}9.4}_{15.8}$  & Chandra HETG & M03\\
\hline
\end{tabular}
\end{center}
$^a${Equivalent width in m\AA.  For those studies listing equivalent 
widths in energy units rather than wavelength units, 
the following conversion was used: $W_\lambda {\rm (m\AA)} = (\lambda^2/hc)
W_E  = 37.5 W_E {\rm (eV)}$.  Errors for all values are quoted are at 90\% confidence, except the C02 value for Mrk\,421, which is a 1$\sigma$ estimate.}
\\
$^b${References: C02 = Cagnoni (2002);
F02 = Fang et al. (2002); F03 = Fang, Sembach, \& Canizares (2003);
M03 = McKernan et al. (2003a);
N02 = Nicastro et al. (2002); R03 = Rasmussen, Kahn, \& Paerels (2003).}
\end{table}

\subsection{An Estimate of X-ray Absorption Produced by the Hot Gas at $z\approx0$}
Perhaps the most direct measure of the amount of hot gas is given by 
the X-ray absorption measurements. A tenuous hot Galactic corona or Local 
Group gas should be revealed through X-ray absorption-line observations of 
O\,{\sc vii}. The medium should have a temperature $T \ga 10^6$\,K to 
avoid direct detection in lower ionization species such as O\,{\sc vi},
a density $n_H \la 10^{-4}$ cm$^{-2}$ to prevent orbital decay of the Magellanic 
Stream, and an extent $L \ga 70$ kpc to explain the detections of O\,{\sc vi}
in the Magellanic Stream (see Sembach et al. 2003).
The column density of O\,{\sc vii} in the hot gas
is given by
\begin{center}
N(O\,{\sc vii}) = (O/H)$_\odot$~$Z/Z_\odot$~f$_{\rm O\,VII}$~$n_H$~$L$,
\end{center}
\noindent
where $Z/Z_\odot$ is the 
metallicity of the gas in solar units, 
f$_{\rm O\,VII}$ is the ionization fraction
of O\,{\sc vii}, $L$ is the 
path length, and (O/H)$_\odot = 4.90\times10^{-4}$ (Allende Prieto, Lambert, \& Asplund
2001).  
At $T \sim 10^6$\,K,  f$_{\rm O\,VII} \approx 1$ 
(Sutherland \& Dopita 1993). For $n_H = 10^{-4}$ cm$^{-3}$,  
$N$(O\,{\sc vii}) $\sim 1.5\times10^{16}~Z/Z_\odot$~($L / 100$ kpc) cm$^{-2}$.
This column density is of the order of magnitude derived for the equivalent 
widths listed in Table~3.  Conversion of the observed O\,{\sc vii} equivalent 
widths into an O\,{\sc vii} column densities requires assumptions about 
the Doppler parameters for the lines since the lines are likely unresolved, 
with some authors preferring to constrain
the possible b-values  (e.g., Fang et al. 2003; Nicastro et al. 2002), and 
others preferring to list only lower limits to N(O\,{\sc vii}) from the assumption
of a linear curve of growth (e.g., Rasmussen et al. 2003). 

\begin{figure}
\includegraphics{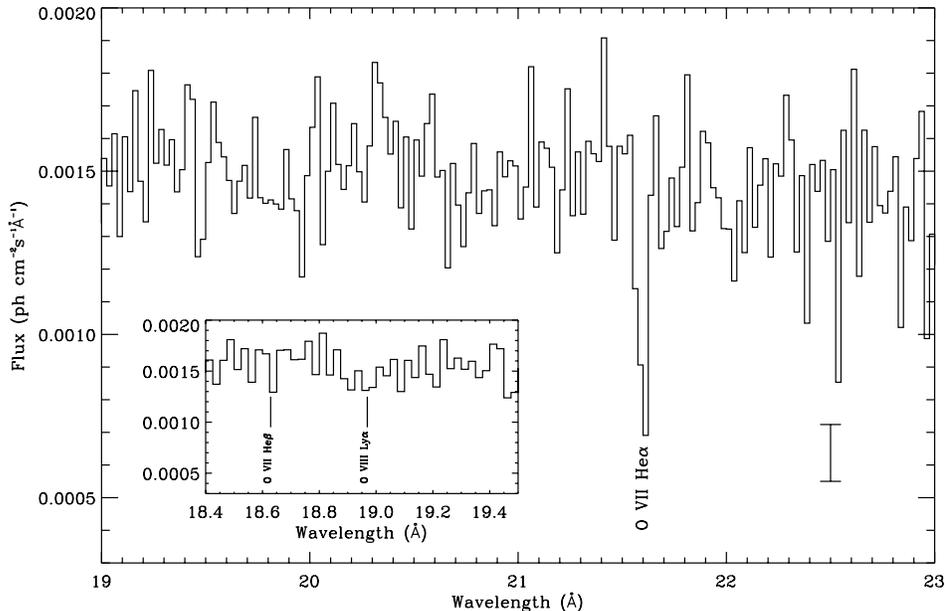}
\vspace{3.3in}
\caption{\small
A portion of the Chandra LETG-ACIS spectrum of 3C\,273 showing a well-detected
zero-redshift O\,{\sc vii} He$\alpha$ line at 21.6 \AA, with $W_\lambda =
28.4\pm^{12.5}_{6.3}$\,m\AA\ at 90\% confidence (Fang et al. 2003).  The 
O\,{\sc vii} may arise in hot gas within the Milky Way or within a hot 
Local Group medium.  The error bar in the lower right corner indicates the 
typical $1\sigma$ photon-counting error on each data point. }
\end{figure}

It is extremely 
important to realize that there may be sources for the O\,{\sc vii}
and other high ionization X-ray lines observed that do not involve a hot Galactic 
corona or Local Group medium.  Alternative locations for some of the 
hot gas along the sight lines
observed include the thick disk and low halo of the Galaxy 
(see Savage et al. 2003)
and large Galactic structures  that are filled with hot gas, such as Loop~I
(see Snowden et al. 1997).  The possible contributions of these 
different
regions to the observed O\,{\sc vii} absorption along the 3C\,273 sight line
are discussed by Fang et al. (2003); a portion of their X-ray spectrum is reproduced
in Figure~2.  
Isolating the distant Galactic corona or Local Group contributions 
to the X-ray absorption from nearby regions of hot gas
is difficult to do kinematically because the spectral resolution of the X-ray 
data is insufficient to resolve the velocities of most objects in the Local Group
from the velocity of the Galaxy. 

A hot extended Galactic corona or Local Group medium has several testable 
predictions.  O\,{\sc vii} absorption should be observed in essentially any 
direction where an X-ray bright  background continuum source can be observed 
with sufficient signal-to-noise to detect the $\lambda21.60$ line with a strength
$W_\lambda \ga 15$ m\AA.  The amount of absorption observed may vary based on
direction and the types of foreground structures probed.  In general, 
observations of nearby X-ray sources should yield 
interstellar O\,{\sc vii} column densities less than those observed toward 
extragalactic X-ray sources.  Futamoto et al. (2003) have recently 
reported
the detection of strong O\,{\sc vii} absorption toward the low mass X-ray 
binary 4U1820-303 in NGC\,6624, which is at odds with this prediction 
since 4U1820-303 is only $\sim1$ kpc from the 
Galactic plane.  However, Futamoto et al. (2003) did not detect O\,{\sc vii} 
toward Cyg-X2, which is at a similar Galactic altitude.  
Additional observations of this type would be valuable 
for determining the distribution of hot gas near the plane of the Galaxy.
Investigators working in this field should 
report significant non-detections whenever possible.

\subsection{Problems with a Single-Phase Ultra-Low Density Medium Model}

Given the probability that several regions 
contribute to the observed absorption, it is dangerous to assume that the X-ray bearing
gas is uniform, at a single temperature, or confined to one type of region.  The same 
holds true for the 
lower temperature O\,{\sc vi} gas, which is known to have a patchy 
spatial distribution 
and line profiles that contain multiple components (see Savage et al. 2003; Sembach
et al. 2003; Howk et al. 2002).  An example of the pitfalls that can be encountered
 in analyses of the X-ray absorption  is provided by the PKS\,2155-304 sight line.    
Nicastro et al. (2002) associated the  X-ray 
absorption near $z \sim 0$ along the PKS\,2155-304 sight line with the high-velocity 
O\,{\sc vi} absorption observed by Sembach et al. (2000).  They found that  a relatively
uniform, ultra-low density, single-phase plasma with a density 
$n_e \sim 6\times10^{-6}$ cm$^{-3}$ and a size of roughly 3 Mpc was able to explain 
the O\,{\sc vi-viii} absorption, modulo some difficulties in reproducing 
the observed Ne/O 
ratio in the hot gas.  From this, they concluded that the X-ray bearing 
gas must be distributed throughout the Local Group.  However, this conclusion rested strongly
on the supposition that the high velocity O\,{\sc vi} absorption and the X-ray 
absorption in this direction are uniquely related, with no other significant sources of 
O\,{\sc vi} or O\,{\sc vii-viii} inside or 
outside the Galaxy.  This conclusion fails to account for the complex distribution 
of gas along the sight line and the ample evidence that the O\,{\sc vi} HVCs 
in this direction have kinematical signatures similar to lower ionization species
whose presence is not consistent with gas at $T = (0.5-1.0)\times10^6$\,K and 
$n_e \sim 10^{-6}$ cm$^{-3}$ [see Sembach 
et al. (1999) and Collins et al. 
(2004) for information on the lower ionization species
such as C\,{\sc ii}, C\,{\sc iv}, Si\,{\sc ii}, Si\,{\sc iii}, Si\,{\sc iv}].
The ionization parameter in the proposed 
medium is far too high to explain the strengths of the low ionization lines 
observed at essentially the same velocities as the O\,{\sc vi} HVCs.   
Other types of inhomogeneities in the hot gas distribution may also be 
present, as they are needed to reconcile the
observed Ne/O ratio with a solar ratio of Ne/O (Nicastro et al. 2002).
(The possibility of explaining the factor of 2 discrepancy between the 
observed Ne/O ratio and the solar ratio by preferential incorporation of 
oxygen into dust grains 
seems dubious given the much higher cosmic abundance of oxygen).
If one invokes inhomogeneities in the gas distribution to explain
the lower ionization gas, then such regions must necessarily 
account for some, or perhaps even all, of the O\,{\sc vi}. As a result,
the ionization constraints used to deduce the existence of an
 ultra-low density medium break down, and the adoption of a uniform single 
temperature medium is cast in doubt.

The low ionization high velocity clouds toward  PKS\,2155-304 should have
densities typical of those of clouds located near the Galaxy
(i.e., $n \sim 10^{-3}-10^{-1}$ cm$^{-3}$).  An
example of one such region is high velocity cloud Complex~C, whose ionization
conditions suggest that the cloud is interacting with gas in
an extended Galactic corona (Sembach et al. 2003; Fox et al. 2004;
Collins et al. 2004).
Interactions of this type are consistent with the wider distribution of 
high velocity O\,{\sc vi} seen on the sky (see \S4) and provide a dynamic, 
non-equilibrium view of the gas ionization that is far more amenable to 
creating a range of ionization conditions than is possible in 
a static, ultra-low density medium having enormous cooling times.

Thus, while some of 
the highly ionized oxygen may be located in the Local Group,  consideration 
of all of the available information indicates that an ultra-low density 
($n \sim 10^{-6}$ cm$^{-3}$, $\delta \sim 60$) 
Local Group medium is not a viable explanation for all of the X-ray and 
high velocity O\,{\sc vi} absorption toward PKS\,2155-304.    
A low density (but not too low density)
medium with $n \sim 10^{-4}-10^{-5}$ cm$^{-3}$ is still a possible repository
for much of the highly ionized gas, but is in all likelihood
only part of a complex distribution of hot gas along this sight line.
Additional discussion of the observational and theoretical implications of a  
Local Group medium and the complications that can arise in interpreting the existing data can be found in Maloney (2003).

\section{High Velocity Clouds in the Vicinity of the Milky Way}

We have conducted an extensive 
 study of the highly ionized high velocity gas in the 
vicinity of the Milky Way using data from the FUSE satellite.  
Below, I summarize the results for  the sight lines toward 100 AGNs/QSOs 
and two distant halo stars (see Sembach et al. 2003; Wakker et al. 2003).  
The FUSE observations represent the culmination of $\sim 4$ megaseconds
of actual exposure time and many years of effort by the FUSE 
science team.
For the purposes of this study, gas with 
$|v_{LSR}| \ga 100$ km~s$^{-1}$ is typically identified as 
``high velocity'', while lower velocity gas is attributed to the Milky Way 
disk and halo.  Sample spectra from the survey are shown in Figure~3.
Complementary talks on this subject were presented at the Sydney IAU in
Symposium 217 by Blair Savage (a description of the FUSE O\,{\sc vi} survey) 
and Bart Wakker (an overview of high velocity clouds) 
(see Savage et al. 2004; Wakker 2004).

\begin{figure}
\includegraphics{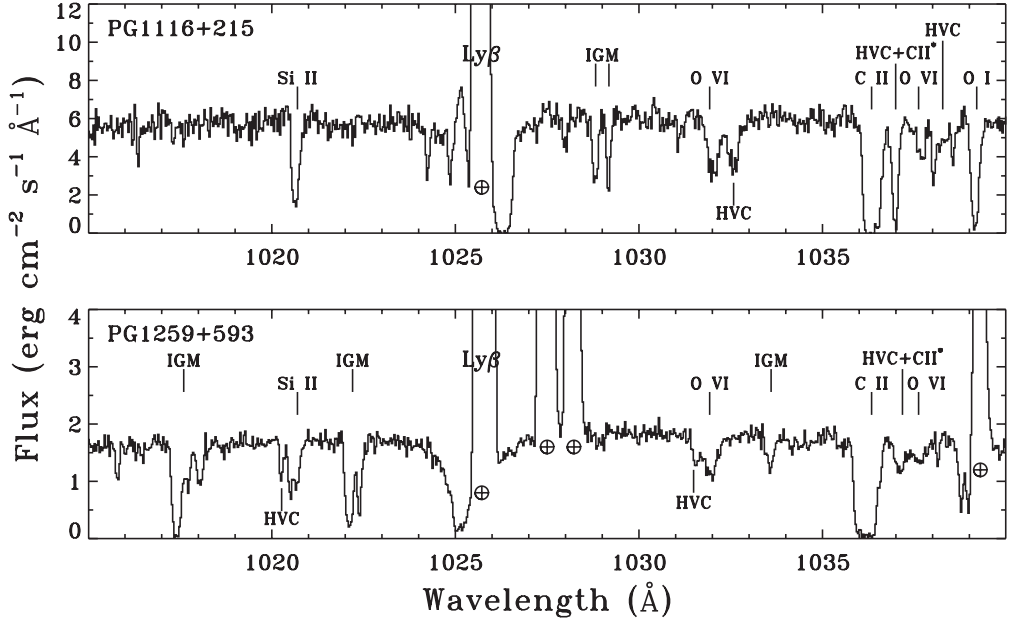}
\vspace{3.4in}
\caption{\small  Observations of two objects in the FUSE high-velocity 
O\,{\sc vi} survey.  The data have 
a velocity resolution of $\sim20$ \kms\ (FWHM) and are binned to $\sim10$ \kms\
($\sim0.033$\,\AA) samples.  Prominent interstellar lines, including the 
two lines of the O\,{\sc vi} doublet at 1031.926\,\AA\ and 1037.617\,\AA,
are identified above each spectrum at their rest wavelengths.  
High velocity O\,{\sc vi} absorption is present along both sight lines.
H\,{\sc i} and metal lines from 
intervening intergalactic clouds are marked in both panels.  Unmarked absorption
features are interstellar H$_2$ lines.
Crossed circles mark the locations of terrestrial airglow lines of H\,{\sc i}
and O\,{\sc i}. From Sembach et al. (2003).
}
\end{figure}

\subsection{Detections of High Velocity O\,{\sc vi}}
Sembach et al.\ (2003) identified 84 individual 
high velocity O\,{\sc vi} features along 102 sight lines observed by FUSE. 
A critical part of this identification process involved detailed consideration
of the absorption produced by O\,{\sc vi} and other species 
(primarily H$_2$) in the thick disk and halo of the Galaxy, as well as the 
absorption produced by low-redshift intergalactic absorption lines of 
H\,{\sc i} and ionized metal species along the lines of sight studied.
Our methodology for identifying the high velocity features and the possible 
complications involved in these identifications
are described in detail by Wakker et al.\ (2003).
We searched for absorption in an LSR velocity 
range of $\pm1200$ km~s$^{-1}$ centered on the O\,{\sc vi} 
$\lambda1031.926$ line.  With few exceptions, the high velocity O\,{\sc vi}
absorption is confined to $100 \le |v_{LSR}| \le 400$ km~s$^{-1}$,  
indicating that the identified O\,{\sc vi} features are either associated with 
the Milky Way or are nearby clouds within the Local Group.  Information
about the lower velocity ($|v_{LSR}| \le 100$ \kms) gas inside the thick
disk and low halo of the Milky Way can be found in Savage et al.\ (2003).

We detect high velocity O\,{\sc vi} $\lambda1031.926$ absorption 
with total equivalent widths $W_\lambda > 30$ m\AA\ at 
$\ge 3\sigma$ 
confidence along 59 of the 102 sight lines surveyed.  For the highest 
quality sub-sample of the dataset, the high velocity detection frequency 
increases to 22 of 26 sight lines.  Forty of the 59 
sight lines have  high velocity O\,{\sc vi} $\lambda1031.926$ absorption
with  $W_\lambda > 100$ m\AA, and 27 have total equivalent widths 
$W_\lambda > 150$~m\AA.  
Converting these O\,{\sc vi} equivalent width detection frequencies
 into estimates of $N$(H$^+$) in the 
hot gas indicates
that $\sim60$\% of the sky (and perhaps as much as $\sim85$\%) 
is covered by hot ionized hydrogen at a level of 
$N({\rm H}^+)  \stackrel{_>}{_\sim} 10^{18}$ cm$^{-2}$,
assuming an ionization fraction f$_{\rm O\,VI} < 0.2$ and a gas
 metallicity similar to that of the Magellanic Stream ($Z/Z_\odot 
\sim0.2-0.3$). This detection frequency of 
hot H$^+$ associated with the  high velocity O\,{\sc vi} is
 larger than the value of $\sim37$\% found for warm, high velocity neutral 
gas with 
$N$(H\,{\sc i})\,$\sim 10^{18}$ cm$^{-2}$ traced through 21\,cm emission
(Lockman et al. 2002).

\subsection{Velocities}

The high velocity O\,{\sc vi} features have velocity centroids ranging 
from $-372 < v_{LSR} < -90$ km~s$^{-1}$ to 
$+93 < v_{LSR}
< +385$ km~s$^{-1}$. There are an additional 6  
confirmed or very likely ($>90$\% confidence) detections and 2 tentative 
detections of O\,{\sc vi} 
between $v_{LSR} = +500$ and +1200 km~s$^{-1}$; these very high velocity
features probably trace intergalactic gas beyond the Local Group. 
Most of the high velocity O\,{\sc vi} features have velocities incompatible 
with those of Galactic rotation (by definition).  

\begin{figure}
\includegraphics{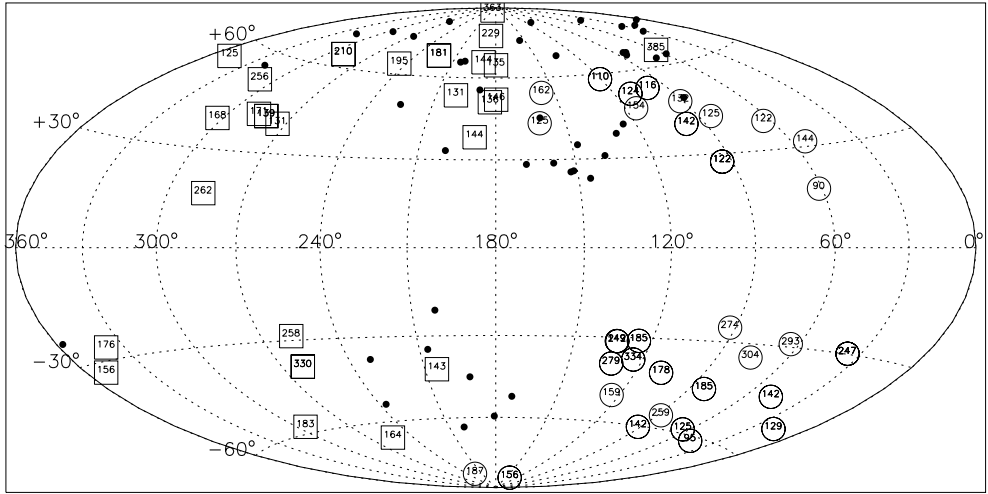}
\vspace{2.9in}
\caption{\small All-sky Hammer-Aitoff 
projection of the high velocity O\,{\sc vi} 
features along the sight lines in the FUSE O\,{\sc vi} survey.  Squares
indicate positive velocity features.  Circles indicate negative 
velocity features.  The magnitude of the velocity is given inside each circle
or square for the highest column density O\,{\sc vi} HVC component along 
each sight line.  Solid dots indicate sight lines with no high velocity 
O\,{\sc vi} detections.}
\end{figure}

We plot the locations of the high velocity O\,{\sc vi} features in 
Figure~4, with squares indicating positive velocity features and circles 
indicating negative velocity features.  The magnitude of the LSR velocity 
(without sign) is given inside each square or circle for the dominant
high velocity O\,{\sc vi} feature along each sight line.  For sight lines
with multiple high velocity features, we have plotted the information 
for the feature with the greatest O\,{\sc vi} column density.   
Solid dots indicate directions where no high velocity O\,{\sc vi} was detected.

There is sometimes good correspondence between high velocity 
H\,{\sc i} 21\,cm emission and high velocity O\,{\sc vi} absorption.
When high velocity H\,{\sc i} 21\,cm emission
is detectable in a particular direction,
high velocity O\,{\sc vi} absorption is usually detected if a suitable 
extragalactic continuum source is bright enough for FUSE to observe.   
For example, 
O\,{\sc vi} is present in the Magellanic Stream, which passes 
through the south Galactic pole and extends up to $b \sim -30^o$,
with positive velocities for $l \ga 180^o$ and 
negative velocities for $l \la 180^o$.
O\,{\sc vi} is present in high velocity cloud Complex~C, which 
covers a large portion of the 
northern Galactic sky between $l=30^o$ and $l = 150^o$ and has 
velocities of roughly --100 to --170 \kms.
The H\,1821+643 sight line ($l=94.0^o, b=27.4^o$) contains O\,{\sc vi} 
absorption at the 
velocities of the Outer Arm ($v \sim -90$ \kms) 
as well as at more negative velocities.  

In some directions, high velocity O\,{\sc vi} is observed with no
corresponding high velocity H\,{\sc i} 21\,cm emission.  For example,
at $l \sim 180^o, b > 0^o$ there are many O\,{\sc vi} HVCs 
with $\bar{v} \sim +150$ \kms.  Some of these features are broad 
absorption wings extending from the lower velocity absorption produced by
the Galactic thick disk/halo.  
High velocity O\,{\sc vi} features toward  Mrk\,478 ($l=59.2^o,
b=+65.0^o$, $\bar{v} \approx +385$ \kms), 
NGC\,4670 ($l=212.7^o, b = +88.6^o$, $\bar{v} \approx +363$ \kms), 
and Ton\,S180 
($l=139.0^o, b=-85.1^o$, $\bar{v} \approx +251$ \kms) stand out as 
having particularly unusual velocities compared to those of other 
O\,{\sc vi} features in similar regions 
of the sky. They too lack counterparts in H\,{\sc i} 21\,cm emission.
These features  may be 
located outside the Local Group (i.e., in the IGM).  We are currently
investigating the H\,{\sc i} content of some of these clouds through
their H\,{\sc i} Lyman-series absorption in the FUSE data.

The segregation of positive and negative velocities in Figure~4 is striking, 
indicating that the clouds and the underlying (rotating) disk of the Galaxy 
have very different kinematics.  A similar velocity pattern is seen for 
high velocity H\,{\sc i} 21\,cm emission and has sometimes been used to 
argue for a Local Group location for the high velocity clouds (see Blitz
et al. 1999).  The kinematics of the high velocity
O\,{\sc vi} clouds are consistent with a distant location, but do not 
necessarily require an extended Local Group distribution as proposed 
by  Nicastro
et al. (2003). The dispersion about the 
mean of the high velocity O\,{\sc vi}
centroids  decreases when the velocities are converted from the
Local Standard of Rest (LSR) into the Galactic Standard of Rest (GSR) and 
the Local Group Standard of Rest (LGSR) reference frames.   While this 
reduction is expected if the 
O\,{\sc vi} is associated with gas in a highly extended Galactic corona or 
in the Local Group,  it {\it does not} provide sufficient proof by itself of an
extended extragalactic distribution for the high velocity gas because the
correction to the LGSR reference frame requires proper knowledge of the total
space velocities of the clouds.  Only one component of motion - the 
velocity toward the Sun - is observed.  Additional information,
such as the gas metallicity, ionization state, or parallax resulting
from transverse
motion across the line of sight, is needed to constrain 
whether the clouds are located near the Galaxy or are farther away in the 
Local Group.

\subsection{Column Densities and Line Widths}
The high velocity O\,{\sc vi} features have logarithmic column densities
(cm$^{-2}$)
of 13.06 to 14.59, with an average of $\langle \log N \rangle = 
13.95\pm0.34$ and a median of 13.97 (see Figure~5, left panel).   
The average high velocity O\,{\sc vi} column density is a factor of 
2.7 times lower 
than the typical low velocity O\,{\sc vi} column density found for the same
sight lines 
through the thick disk/halo of the Galaxy (see Savage et al. 2003).

\begin{figure}
\includegraphics{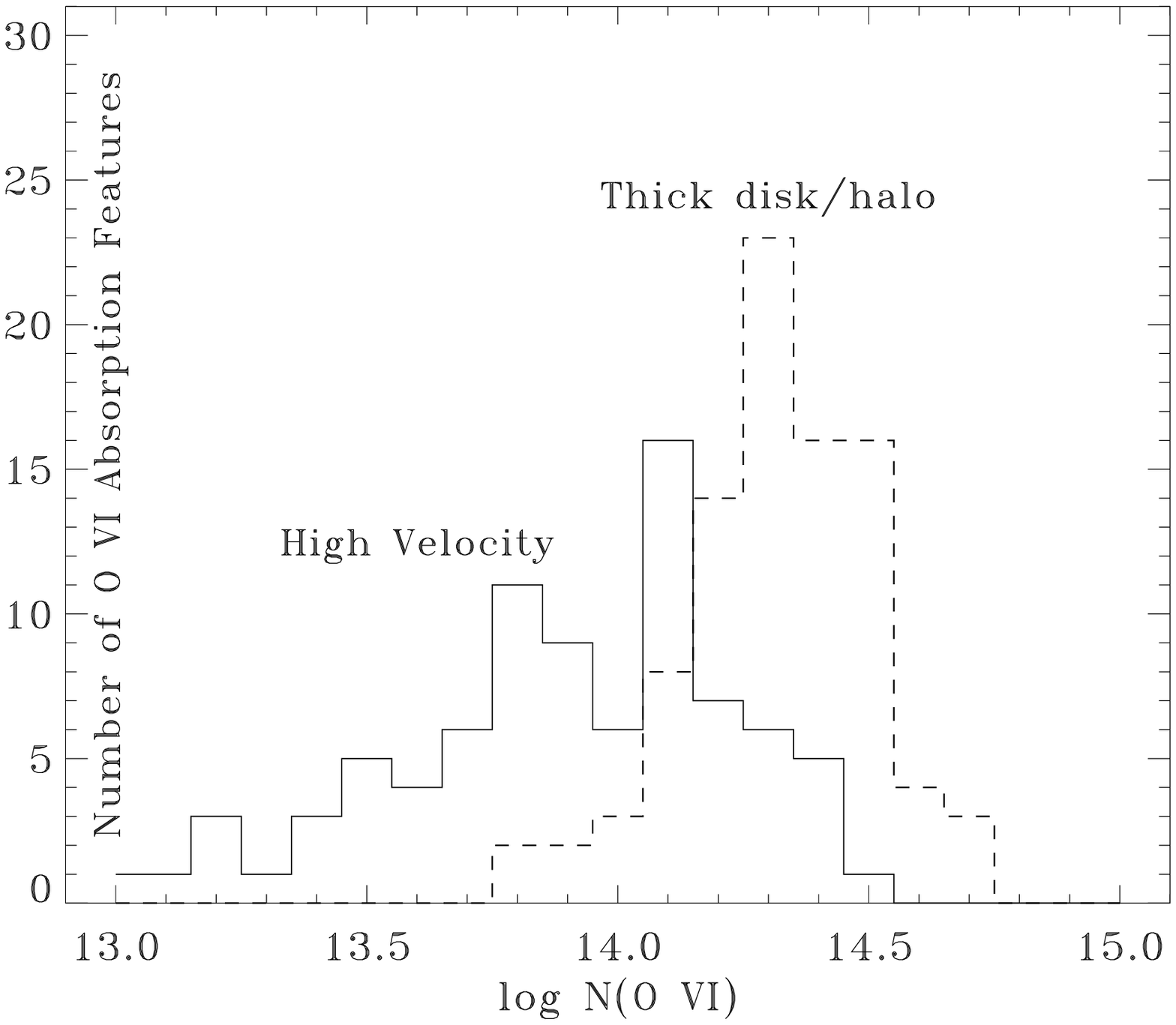}
\includegraphics{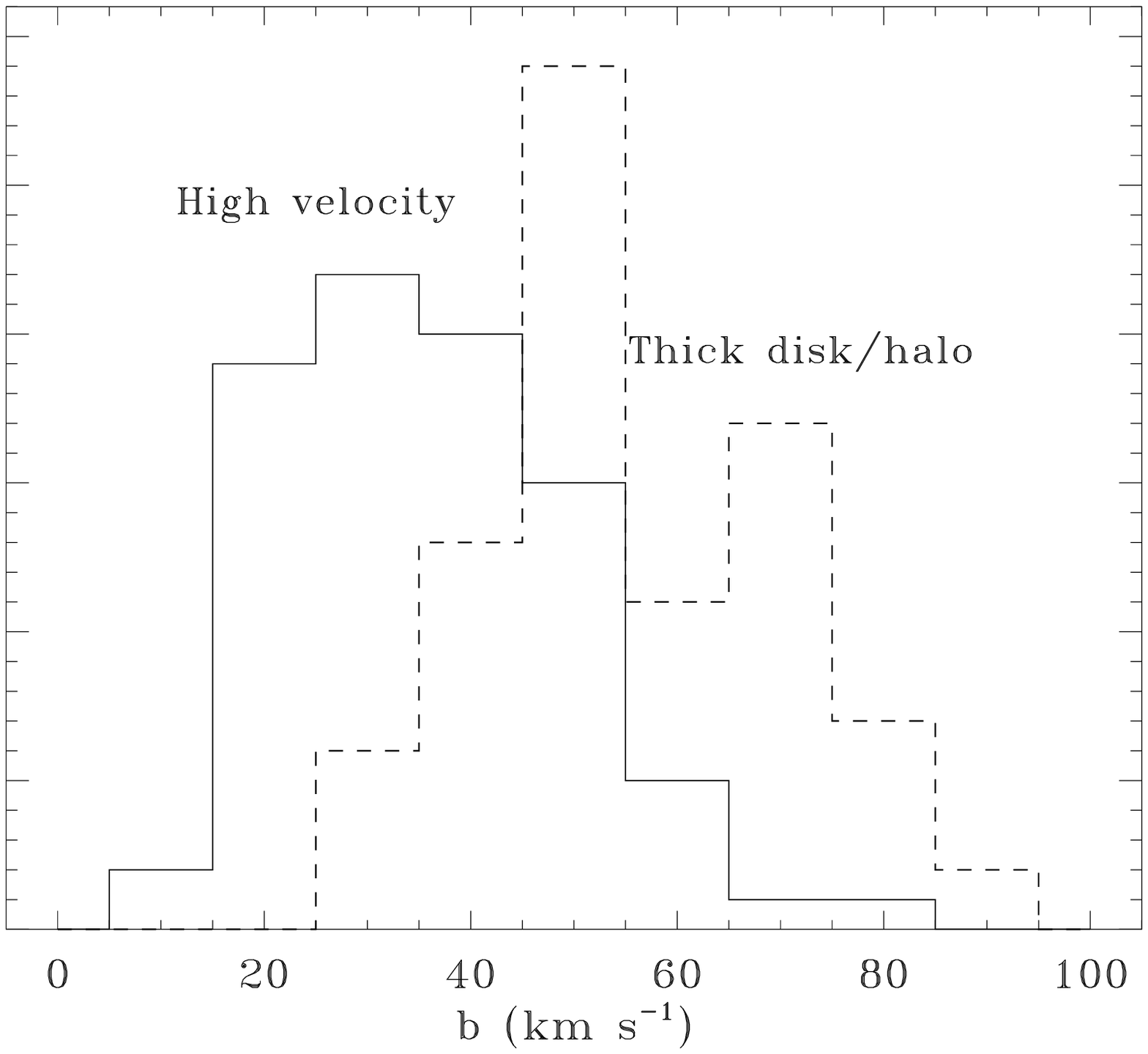}
\vspace{2.5in}
\caption{\small
Histograms of the high velocity O\,{\sc vi} column densities and line
widths (solid lines). The bin sizes are 0.10 dex and 10 \kms, respectively.
For comparison, the distributions for the O\,{\sc vi} absorption
arising in the thick disk and halo of the Galaxy are also shown 
(dashed lines) (from Sembach et al. 2003).}
\end{figure}

The line widths of the high velocity O\,{\sc vi} features range from
$\sim$16 km~s$^{-1}$ to $\sim$81 km~s$^{-1}$, with an average of 
$\langle {\rm b} \rangle = 40\pm14$ km~s$^{-1}$ (see Figure~5, right panel).
The lowest values of b are close to the thermal width of 17.1 km~s$^{-1}$
expected for O\,{\sc vi}
at its peak ionization fraction temperature of $T = 2.8\times10^5$\,K in
collisional
ionization equilibrium (Sutherland \& Dopita 1993).
The higher values of b 
require additional non-thermal broadening mechanisms or gas
temperatures significantly larger than $2.8\times10^5$\,K.

\subsection{Origin of the High Velocity O\,{\sc vi}}

One possible explanation for some of the high velocity O\,{\sc vi} is that
transition temperature gas arises at the boundaries between cool/warm 
clouds of gas and a very hot ($T > 10^6$\,K) Galactic corona or Local Group
medium.   Sources of the high velocity material might include infalling or 
tidally disturbed galaxies.  
 A hot, highly extended ($R > 70$ kpc)
corona or Local Group medium might be left over from the formation of the 
Milky Way or Local Group, or may be the result of continuous accretion of 
smaller galaxies over time.  Evidence for a hot Galactic corona or 
Local Group medium is given in \S3.
Hydrodynamical simulations of clouds moving through a hot, low-density 
medium show that weak bow shocks develop on the leading edges of the 
clouds as the gas is compressed and heated (Quilis \& Moore 2001).  
Even if the clouds are 
not moving at 
supersonic speeds relative to the ambient medium, some viscous or turbulent
stripping of the cooler gas likely occurs.
An alternative explanation for the O\,{\sc vi} observed at high velocities 
may be that the clouds and any associated H\,{\sc i} fragments are simply 
condensations within
large gas structures falling onto the Galaxy.
Cosmological structure formation models predict large numbers of cooling 
fragments embedded in dark matter, and some of these structures should be 
observable in O\,{\sc vi} absorption as the gas passes through the 
$T=10^5-10^6$\,K
temperature regime  (Dav\'e et al. 2001). 

\subsection{Are the O\,{\sc vi} HVCs Extragalactic Clouds?}

Some of the high velocity O\,{\sc vi} clouds may be extragalactic 
clouds, based on what we currently know about their ionization 
properties.  However, claims that
essentially  {\it all} of the O\,{\sc vi} HVCs are extragalactic 
entities associated with an extended Local Group filament based on
kinematical arguments alone appear to be untenable.
 Such arguments
fail to consider the selection biases inherent in the O\,{\sc vi} 
sample, the presence of neutral (H\,{\sc i}) and lower ionization 
(Si\,{\sc iv}, C\,{\sc iv})
gas associated with some of the O\,{\sc vi} HVCs, and the known ``nearby''
locations for at least two of the primary high velocity complexes 
in the sample ---  the Magellanic Stream is circumgalactic tidal debris, and 
Complex~C is interacting with the Galactic corona (see Fox et al.
2004).  Furthermore, the  
O\,{\sc vii} X-ray absorption measures used to support an extragalactic 
location have not yet been convincingly tied to either the O\,{\sc vi} HVCs
or to a Local Group location.  The O\,{\sc vii} absorption may well have a 
significant Galactic component in some directions
(see Fang et al. 2003). The Local Group filament 
interpretation (Nicastro et al. 2003) may be suitable for some of the 
observed high velocity O\,{\sc vi} features, but it clearly fails in 
other particular cases (e.g., the Magellanic Stream or the PKS\,2155-304 
HVCs -- see
\S3).  It also does not reproduce some of the properties of the
ensemble of high velocity O\,{\sc vi} features
in our sample.  For example, the ``Local Supercluster Filament'' model
(Kravtsov, Klypin, \& Hoffman  2002) predicts average O\,{\sc vi}
velocity centroids higher than those 
observed ($\langle \bar{v} \rangle \sim 1000$ 
\kms\ vs. $\langle \bar{v} \rangle < 400$ \kms) and average O\,{\sc vi}
line widths 
higher than
those observed (FWHM $\sim 100-400$ \kms\ vs. FWHM $\sim 30-120$ \kms).
Additional absorption and emission-line
observations of other ions at ultraviolet wavelengths, particularly
C\,{\sc iv} and Si\,{\sc iv}, would provide
valuable information about the physical conditions, ionization, and 
locations of the O\,{\sc vi} clouds.

\section{Summary and Future Prospects}
The detection of hot gas locally and at low redshift has
altered our perspective on the presence of highly ionized
gas outside of galaxies and has led to refined  estimates of the baryonic 
content of the present-day universe.  Clearly, there is much work yet
to be done in determining the spatial distribution of the hot gas, its 
physical conditions, and its association with galaxies and the larger-scale
gaseous structures from which galaxies form.  

A deeper understanding of the spatial distribution of the hot gas 
would be possible with ultra-sensitive maps of the diffuse O\,{\sc vi}
emission associated with the cosmic web.  Emission maps would enable estimates
of the filling factor of the hot gas, and would set the context for ongoing 
absorption-line studies of the hot gas.  Experiments to produce O\,{\sc vi}
emission maps over large regions of the sky are well-suited to small and 
medium Explorer-class missions and have been proposed to NASA. 

The Hubble Space Telescope Cosmic Origins 
Spectrograph, which is to be installed as part of the next HST servicing 
mission, will greatly enhance studies of O\,{\sc vi} systems
at redshifts $0.14 <  z <0.5$ by increasing the number of 
accessible background sources to use for the absorption-line measurements.
Larger redshift paths will be probed, and more sight lines will be 
examined.  In addition,  the high sensitivity of the spectrograph will
 permit the 
acquisition of very high signal-to-noise spectra for bright 
extragalactic sources.
In the longer term, larger (6-20\,m) ultraviolet/optical telescopes
in space will make it possible to select many closely spaced lines of sight 
for detailed spectroscopic and imaging investigations of galactic and 
intergalactic structures associated with the hot gas.  A key scientific 
motivation for such investigations will be to determine  
feedback mechanisms operating on the hot gas and the degree to which these 
processes affect the formation and evolution of galaxies.

\acknowledgments
I thank Blair Savage, Bart Wakker, Philipp Richter, and Marilyn Meade 
for their efforts in making the FUSE O\,{\sc vi} survey 
a reality.  I acknowledge lively discussions about O\,{\sc vi}
absorption with Todd Tripp and Mike Shull.

\end{document}